\documentclass[journal]{IEEEtran}
\AtBeginDocument{%
  }

\usepackage{amsmath}
\usepackage{algorithm}
\usepackage{algpseudocode}
\usepackage{booktabs}
\usepackage{graphicx} 
\usepackage{pifont}
\usepackage{listings}
\usepackage{xcolor}
\usepackage{multirow}
\usepackage{url} 
\usepackage{array}

\newcommand{\cmark}{\textcolor{green}{\ding{51}}}
\newcommand{\xmark}{\textcolor{red}{\ding{55}}}
\usepackage{makecell}

\algrenewcommand\algorithmicrequire{\textbf{Input:}}
\algrenewcommand\algorithmicensure{\textbf{Output:}}

\begin{document}

\title{VeriPilot: An LLM-Powered Verilog Debugging Framework}

\author{Yihan Wang, 
Cheng Liu,~\IEEEmembership{Senior Member,~IEEE,} 
Jiazheng Zhang, 
Lei Zhang,
Long Cheng,~\IEEEmembership{Senior Member,~IEEE,} 
Xiaowei Li,~\IEEEmembership{Senior Member,~IEEE,} 
and Huawei Li,~\IEEEmembership{Senior Member,~IEEE}
\thanks{The corresponding author is Cheng Liu.}
\thanks{This work is in part supported by the Strategic Priority Research Program of the Chinese Academy of Sciences under Grant No. XDB0660103, and the National Natural Science Foundation of China (NSFC) under Grant No. 62174162.}
\thanks{Yihan Wang, Cheng Liu, Jiazheng Zhang, Huawei Li, and Xiaowei Li are with the State Key Lab of Processors, Institute of Computing Technology, Chinese Academy of Sciences, and the University of Chinese Academy of Sciences, Beijing, China. E-mail: \{wangyihan24s, liucheng, zhangjiazheng24s, zlei, lxw, lihuawei\}@ict.ac.cn.}
\thanks{Long Cheng is with with North China Electric Power University, Beijing 102206, China.
E-mail: lcheng@ncepu.edu.cn.}

}

\maketitle

\begin{abstract}
Verilog debugging remains one of the most time-consuming stages in digital circuit design. Recent advances in Large Language Models (LLMs) have enabled automated debugging; however, most existing approaches rely solely on test outputs and compiler feedback in an end-to-end manner, limiting their effectiveness on complex bugs. A key challenge is that the root cause of an error may be far removed from its observable outputs, making it difficult for LLMs to trace long dependency chains in code. This challenge is further exacerbated in large codebases, where long context lengths hinder efficient reasoning.
To address these limitations, we propose VeriPilot, an LLM-powered debugging framework that leverages golden reference models to enable fine-grained bug localization and repair. VeriPilot goes beyond output-level comparison by aligning internal variable semantics between the Verilog design and its corresponding golden model through LLM-based analysis. It then performs step-by-step signal tracing using Control-Data-Flow Graphs (CDFGs) derived from static analysis, identifying a minimal set of suspicious code regions along with their correct counterparts from the golden model.
These structured insights are subsequently provided to the LLM to guide reasoning and automated code repair. Experimental results on the Comprehensive Verilog Design Problems (CVDP) benchmark from NVIDIA demonstrate that VeriPilot improves the repair success rate of GPT-4o from 54.3\% to 85.71\%, significantly enhancing both bug localization accuracy and repair effectiveness for complex Verilog designs. The source code and benchmark are publicly available at Github \url{https://github.com/YihanWn/VeriPilot.git}.
\end{abstract}


\section{Introduction}
Debugging is a fundamental component of digital circuit design, traditionally ensuring the correctness of manually developed hardware modules. As large language models (LLMs) become increasingly involved in hardware design~\cite{thakur2024verigen, wan2024software, zhao2024codev, liu2024craftrtl, zhang2024mg, tang2025hivegen, gong2025largesmalltransferringcuda}, debugging becomes even more critical. The inherent uncertainty and occasional hallucination in LLM-generated Verilog code necessitate automated debugging mechanisms to ensure reliability and continuously improve generation quality~\cite{liu2023verilogeval, wang2025large, ma2024verilogreader, jha2025large}.

Prior efforts predominantly rely on compilation errors or mismatches in output waveforms as feedback for LLMs to iteratively repair code in an end-to-end manner~\cite{thakur2023autochip, tsai2024rtlfixer, huang2024towards}. While such approaches can handle simple circuits, their effectiveness quickly degrades as design complexity increases~\cite{huang2024towards}. This limitation stems from two fundamental challenges. On the one hand, as the codebase grows, the input context can easily exceed or saturate the effective reasoning capacity of LLMs, leading to degraded performance due to long context lengths and irrelevant information. On the other hand, debugging RTL designs frequently requires tracing long-range dependencies, where the root cause of an error may be far removed from the observed output mismatch. Such long-chain reasoning is inherently difficult for LLMs to perform reliably in an end-to-end fashion. Basically, LLMs often lack precise bug localization, making it difficult to identify where the design deviates from the intended behavior~\cite{xu2024meic, wang2025veridebug, yao2025location}. More critically, existing methods lack explicit references to correct internal behavior. While high-level specifications or golden software models describe the intended functionality, they abstract away low-level hardware details such as intermediate signals, pipeline states, and control logic. As a result, LLMs struggle not only to determine what is wrong, but also what the correct implementation should look like at the RTL level.

\begin{figure}[htbp]
\centerline{\includegraphics[width=1\linewidth]{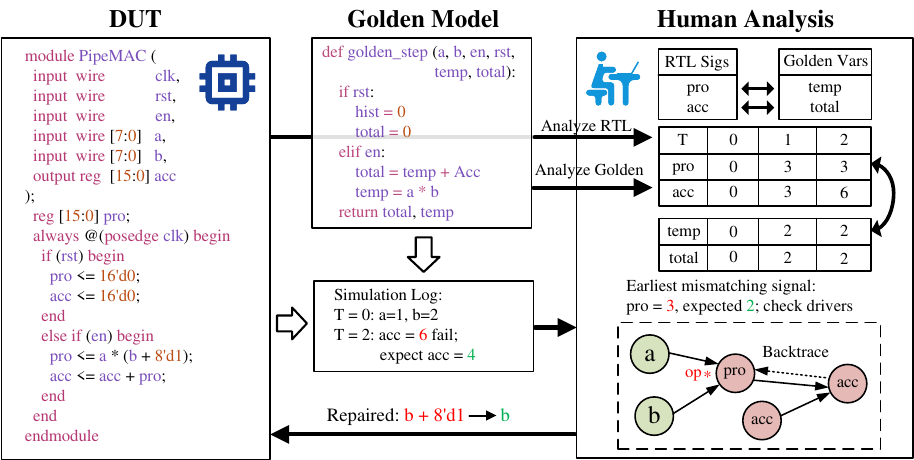}}
\caption{Overview of the manual debugging methodology. The process initiates with an output comparison between the DUT and the golden model to detect failures. By leveraging aligned signals and variables from both entities, engineers then perform a backward traversal through the circuit's internal logic hierarchy to isolate and rectify the root cause of the erroneous behavior.}

\label{HumanDebug-LLMDebug}
\end{figure}

To address these challenges, we revisit the workflow of human hardware debugging (Fig.~\ref{HumanDebug-LLMDebug}). Engineers typically construct a testbench to apply input stimuli to the Design Under Test (DUT) and observe its outputs. When discrepancies arise, they go beyond output inspection by analyzing simulation logs, examining internal signals, and tracing execution paths through the design to identify the root cause~\cite{bergeron2000writing}. This process fundamentally relies on fine-grained observability—particularly the behavior of internal variables—rather than solely on final outputs. A key difficulty, however, lies in the abstraction gap between high-level specifications and RTL implementations. While high-level models describe functional intent using abstract variables, RTL designs introduce numerous low-level signals (e.g., intermediate wires, registers, and control signals) that do not have explicit counterparts in the high-level description. As a result, effective debugging requires identifying meaningful correspondences between high-level variables and low-level hardware signals, which is a task largely depending on human expertise. In practice, engineers iteratively establish and refine such correspondences, progressively shifting the point of comparison from outputs to internal signals along the data path. By aligning semantically related variables across abstraction levels and tracing their dependencies step by step, they can narrow down the discrepancy to a minimal set of suspicious code regions, ultimately enabling precise and fine-grained bug localization and repairing.

Inspired by this human debugging workflow, we propose VeriPilot, an LLM-powered Verilog debugging framework that incorporates golden reference models to enable fine-grained signal-level reasoning. Rather than relying solely on output mismatches, VeriPilot compares both design outputs and internal signal behaviors against a golden model. It aligns variable semantics between the RTL design and the reference model using LLM-guided analysis, and constructs CDFGs via static code analysis. VeriPilot then performs stepwise signal tracing on both CDFGs to identify a minimal set of suspicious code regions, along with their corresponding correct logic from the golden model. These structured insights—localized errors paired with correct references—are provided to the LLM to guide reliable and interpretable code repair. Experimental results on the Strider and CirFix benchmarks demonstrate that VeriPilot significantly improves bug localization accuracy and repair success rates, particularly on complex Verilog designs with deep control and data dependencies. 

In summary, this work makes the following key contributions:
\begin{itemize}
\item We propose a human-inspired debugging paradigm, VeriPilot, the first framework to integrate LLM-based semantic alignment, CDFG-driven trace analysis, and fine-grained reference extraction for Verilog debugging.

\item We present a principled approach for generating structured debugging context, including localized suspicious code segments paired with correct reference logic, substantially improving the LLM’s ability to perform accurate and explainable repairs.

\item We present a comprehensive evaluation on the CVDP benchmark, demonstrating significant gains in bug localization accuracy and repair success on both simple and complex Verilog designs. In particular, VeriPilot improves the repair success rate of GPT-4o from 54.3\% to 77.1\%, showing the effectiveness of our framework in enabling more reliable automated RTL debugging.
\end{itemize}

\begin{figure*}[!t]
    \centering
    \includegraphics[width=1\linewidth]{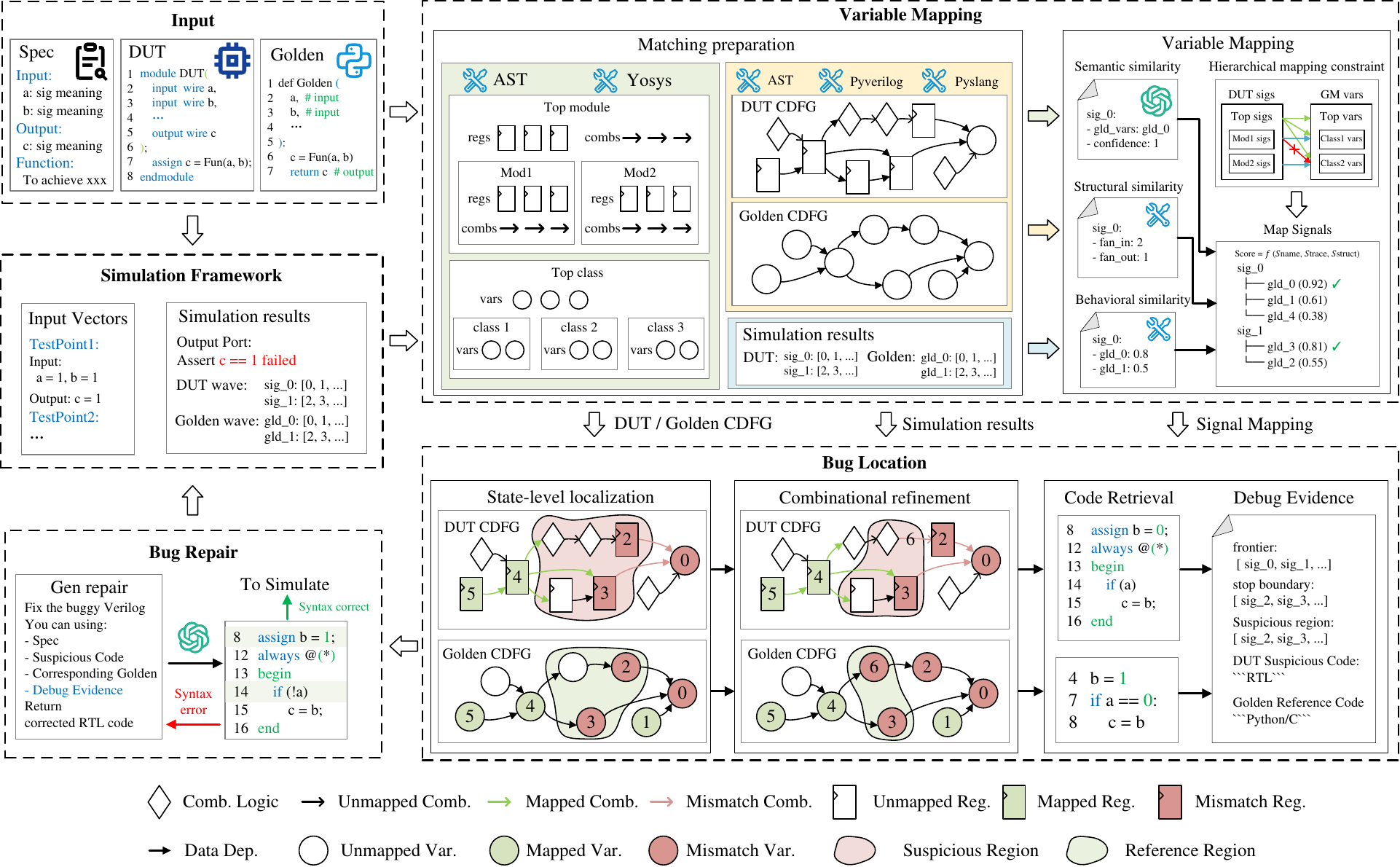}
    \caption{VeriPilot is an LLM-powered multi-agent Verilog debugging framework. It consists of four phases: (1) simulation \& verification to identify the earliest divergent execution state between the DUT and the golden model, (2) variable mapping to align semantically equivalent variables through cross-language CDFGs, (3) bug localization via deterministic backward tracing over aligned dependency graphs, and (4) bug repair using localized debugging evidence to guide LLM-based patch generation. A shared debugging context persistently stores intermediate results and historical debugging information to support iterative multi-round repair.}
    \label{fig:Framework}
\end{figure*}

\section{Related Work}

Debugging is a time-consuming yet essential part of programming. To alleviate this burden, automated program repair (APR) has been extensively studied for software languages~\cite{le2019automated, zhang2023survey, liu2021critical}. Recently, APR techniques have been extended to register-transfer level (RTL) design, though significant challenges remain due to the unique semantics and verification constraints of hardware description languages~\cite{sudakrishnan2008understanding, liu2024craftrtl}. Early RTL-specific APR approaches rely on structural analysis, constraint solving, or heuristic search, often leading to large patch spaces and limited scalability. CirFix~\cite{ahmad2022cirfix} formulates RTL repair as a search problem using simulation-driven scoring and genetic algorithms to explore candidate patches. Strider~\cite{yang2023strider} applies static analysis to localize suspicious regions and repair them using expected signal values; however, it assumes access to ground-truth internal signals, which are typically unavailable in real-world scenarios. RTL-Repair~\cite{laeufer2024rtl} combines template-based patch generation with SMT solving to enforce semantic correctness. While VeriFix~\cite{ahmed2022verifix} focuses on software concurrency bugs, it highlights an important insight for hardware debugging: effective repair requires reasoning about semantic correctness rather than merely satisfying test outputs.

The advent of large language models (LLMs) has opened new avenues for automated RTL debugging~\cite{xu2024llm, qayyum2025llm, link2025evaluating, fang2025lintllm, zuo2025complexvcoder, fu2024generalize}. Existing LLM-based approaches can be broadly categorized into two directions: (1) agentic frameworks that integrate LLMs with compilers and simulators, and (2) model-centric approaches that fine-tune LLMs for end-to-end debugging~\cite{nadimi2025verimind, mi2024coopetitivev, shahidzadeh2024automatic, yao2025hdldebugger, hong2025hdl2v, ahmad2024hardware, pit2021effective, zhao2025vfocus}. Early systems such as AutoChip~\cite{thakur2023autochip} and RTLFixer~\cite{tsai2024rtlfixer} adopt iterative reasoning loops guided by compiler errors or output mismatches. MEIC~\cite{xu2024meic} improves this paradigm via lightweight fine-tuning within a dual-agent architecture, while VeriDebug~\cite{wang2025veridebug} jointly optimizes retrieval and generation for unified bug localization and repair. More recent multi-agent frameworks such as MAGE~\cite{zhao2025mage}, VerilogCoder~\cite{ho2025verilogcoder}, and BugGen~\cite{jasper2025buggen} further decompose debugging into specialized roles and incorporate self-refinement mechanisms.

Despite these advances, most existing methods operate in an end-to-end manner, relying primarily on compiler diagnostics and output-level feedback. Such signals are often too coarse-grained to support precise bug localization, particularly when errors originate far from observable outputs. In addition, RTL debugging typically involves long dependency chains and large codebases, posing challenges for LLMs in both reasoning over extended execution paths and handling long-context inputs. Finally, existing approaches lack explicit alignment between high-level functional intent and low-level hardware implementation, limiting their ability to leverage high-level references for fine-grained debugging. To address these limitations, we propose VeriPilot, a framework that introduces structured, cross-abstraction debugging. Instead of relying solely on end-to-end feedback, VeriPilot leverages LLMs to align high-level variables with RTL signals, compare internal behaviors against a golden reference, and perform stepwise tracing via control-data-flow analysis for precise bug localization. By providing the LLM with both localized error contexts and corresponding correct semantic references, VeriPilot enables more accurate and interpretable RTL repair.

\section{VeriPilot Framework}

A fundamental limitation of prior LLM-based Verilog debugging approaches lies in their reliance on coarse-grained, end-to-end error signals typically limited to simulation mismatches at module outputs or compiler diagnostic messages. While such signals may suffice for simple, localized bugs, they are fundamentally inadequate for complex designs where the root cause of a failure resides several dependency levels upstream from the observable symptom. In such scenarios, the LLM is expected to navigate a long and ambiguous causal chain purely from textual error descriptions, a task that quickly becomes intractable as code complexity grows. Our design philosophy is grounded in a key insight borrowed from established hardware verification practice. Basically, a high-level golden model typically written in a high-level language such as C/C++ or Python encodes the ground-truth behavioral specification of the design. Rather than asking the LLM to infer design intent solely from test failures, we leverage this structured behavioral reference to dramatically narrow the debugging search space. Specifically, by aligning the internal signal semantics of the design under test (DUT) with those of the golden model, we can decompose a complex end-to-end debugging task into a sequence of fine-grained, localized comparisons, each substantially easier for an LLM to reason over. We further observe that systematic backward tracing along program dependency graphs can deterministically identify the earliest divergence point between the DUT and the reference model, transforming opaque black-box debugging into a structured, evidence-driven procedure.

\subsection{Overview}
Instantiating the above philosophy, we propose VeriPilot, an LLM-powered Verilog debugging framework (Fig.~\ref{fig:Framework}). It adopts a high-level golden model as the behavioral specification consistent with standard industrial chip verification practice. The overall system is orchestrated as a multi-agent pipeline in which specialized agents collaborate across four phases including simulation \& verification, variable mapping, bug localization, and bug repair, with structured information passing between phases, persistent memory of historical findings, and tight integration with external static analysis tools. This architectural choice ensures that each agent operates on a well-scoped, information-rich context rather than the full codebase, directly mitigating the long-context reasoning degradation that plagues monolithic LLM-based approaches.

For the simulation \& verification agent, it leverages LLMs to generate a simulation framework according to design inputs such as design specifications, DUT, and the golden reference model. Specifically, it produces identical input vectors for both the golden model and the DUT, and setups the cross-language co-simulation environment. Meanwhile, it handles tool invocation for simulation such as Verilator or VCS for the Verilog DUT and a native interpreter for the golden model. In addition, it also monitors the execution and the comparison of outputs. Whenever output vectors from the two models diverge, it halts the simulation at the earliest divergent cycle and initiates the debugging procedure. The failing input vector, divergent output signals, and the timestep of divergence are captured and stored in a shared debugging context where a persistent structured memory that all subsequent agents can read from and write to.

For the variable mapping agent, we first leverage a suite of static analysis tools to construct Control-Data-Flow Graphs (CDFGs) for both the Verilog DUT and the high-level golden model. Specifically, dedicated CDFG extraction utilities are applied to each design representation, producing deterministic and precise structural artifacts that serve as the foundation for subsequent reasoning. Based on these CDFGs, the agent further employs an LLM to establish cross-language variable correspondences between the DUT and the golden model.
The mapping objective extends beyond aligning output signals. The agent attempts to identify as many semantically equivalent internal variables as possible, enabling fine-grained intermediate error indicators that significantly improve debugging resolution and localization accuracy. To support iterative debugging, previously established mappings, unresolved ambiguities, and intermediate reasoning results are persistently maintained in a shared debugging context and reused across iterations, thereby avoiding redundant recomputation during multi-round debugging sessions. Variable mapping is detailed in Section~\ref{sec:variable-mapping}.

With variable mappings established, the bug localization agent applies a deterministic BackTrace algorithm that traverses aligned variables along the cross-language CDFGs in a backward direction which moves from divergent output signals toward potential root-cause input variables. At each step, the algorithm compares the runtime values of semantically aligned DUT and golden model variables at the failing timestep, propagating suspicion scores toward the nodes with the earliest divergence. This algorithmic traversal is performed as an external tool call, ensuring that the causal tracing logic is precise and computationally grounded rather than hallucinated by the LLM. The result is a minimal suspicious region. It is essentially a small, ranked set of DUT code lines that are most likely responsible for the observed mismatch, each paired with its correct counterpart from the golden model. This structured localization evidence is appended to the shared debugging context for use in the repair phase. The full localization procedure is described in Section~\ref{sec:bug-localization}.

Finally, the Repair Agent aggregates all structured evidence accumulated in the shared debugging context including the suspicious DUT code lines, their golden model counterparts, relevant signal traces, and variable mapping context. Then, it synthesizes a precise repair prompt for the LLM. Leveraging this rich, localized context, the LLM is asked to reason about the design intent and generate a targeted code patch, rather than attempting to comprehend the entire DUT in one pass. The repaired DUT is automatically re-simulated by the Environment Agent, and if the mismatch persists due to multiple independent bugs, the framework initiates a new debugging iteration. Historical repair attempts and their outcomes are retained in the shared context, enabling the Repair Agent to avoid previously explored patches and progressively refine its hypotheses across iterations. This closed-loop, tool-augmented, iteratively refining multi-agent architecture endows VeriPilot with the capability to handle complex, multi-bug Verilog designs that far exceed the practical scope of prior end-to-end LLM debugging approaches.

\subsection{Variable Mapping} \label{sec:variable-mapping}
To enable fine-grained debugging across heterogeneous design representations, we establish variable-level correspondences between the Python golden model and the Verilog RTL. Rather than relying solely on output mismatches, our goal is to align as many semantically equivalent internal variables as possible, thereby exposing intermediate discrepancies that significantly improve error localization accuracy. However, directly matching variables across languages and abstraction levels is challenging due to inconsistent naming conventions, structural transformations, and implementation-specific optimizations. 
In order to address these challenges, we formulate variable alignment as a constrained bipartite matching problem that jointly leverages structural, semantic, and behavioral information. The overall process consists of two stages: (1) \emph{structural representation construction}, which extracts comparable program representations and dependency information from both designs, and (2) \emph{multi-view variable alignment}, which computes cross-language correspondences using complementary similarity metrics under hierarchy-aware constraints.

\subsubsection{Structural Representation Construction} 
We first construct CDFGs for both the Verilog DUT and the Python golden model using dedicated static analysis utilities. The extracted CDFGs provide deterministic and precise structural representations that capture data dependencies, control relationships, and variable interaction patterns in each design. These graph-level artifacts serve as the structural foundation for subsequent variable alignment.

Beyond graph construction, we additionally extract auxiliary metadata for each variable, including module hierarchy information, variable attributes, dependency statistics, and local source-code context. Such information enables the alignment stage to reason about variables not only based on naming patterns, but also according to their structural roles and behavioral characteristics within the overall design.
To further improve scalability, we partition variables into multiple disjoint matching spaces according to hierarchy and structural locality. This decomposition significantly reduces the search space for subsequent matching and avoids unnecessary comparisons between structurally unrelated variables.

\subsubsection{Multi-View Variable Alignment}
After constructing the structural representations, we perform cross-language variable alignment by integrating multiple complementary similarity signals. Relying on a single similarity criterion is often unreliable in practice, especially when design bugs, implementation optimizations, or abstraction differences distort signal values or naming conventions. Therefore, our approach jointly combines semantic, behavioral, and structural similarities to improve matching robustness.

\textbf{Semantic similarity.}
We first estimate the semantic similarity between variables using an LLM-based analysis. Given a golden-model variable $p$ and a DUT variable $v$, we prompt the LLM with their variable names, attributes, and surrounding source-code context to obtain a semantic similarity score $S_{\text{name}}(p, v) \in [0,1]$. Variables with identical or semantically related naming patterns tend to receive higher scores, while unrelated variables receive lower scores.

\textbf{Behavioral similarity.}
To incorporate runtime behavior into the alignment process, we compute behavioral similarity scores $S_{\text{trace}}(p, v)$ using execution traces generated during simulation. The similarity is evaluated according to the consistency of variable value transitions across time. To ensure statistical reliability, behavioral similarity is only computed when the trace length exceeds a minimum threshold $L$; otherwise, this feature is excluded from the final matching score.

\textbf{Structural similarity.}
We further incorporate structural information derived from the constructed CDFGs. For each variable, we characterize its local structural context using lightweight descriptors such as fan-in and fan-out dependency statistics. The structural similarity score $S_{\text{struct}}(p, v)$ is then computed according to the similarity of these descriptors. Unlike exact graph matching, this lightweight representation remains robust under abstraction differences and implementation-specific transformations.

\textbf{Unified matching.}
For every feasible candidate pair $(p,v)$ within the same matching space, we compute a unified similarity score as shown in Equation \eqref{eq:pv_score}.
Using these scores, we construct a weighted bipartite graph in which the left partition represents Python variables and the right partition represents Verilog variables. Each edge $(p,v)$ is weighted by $\mathrm{Score}(p,v)$. The final variable alignment is obtained by solving a maximum-weight bipartite matching problem using the Hungarian algorithm, as shown in Equation \eqref{eq:maximum-weight-bipartite-matching}. This formulation enforces one-to-one correspondences and produces a consistent mapping across the entire matching space. The mappings obtained from all matching spaces are then merged to produce the final cross-language variable mapping $\mathit{Map}$.

\begin{equation}
\label{eq:pv_score}
\begin{aligned}
\mathrm{Score}(p,v)
&= f\!\left(
S_{\mathrm{name}}(p,v),\;
S_{\mathrm{trace}}(p,v),\;
S_{\mathrm{struct}}(p,v)
\right), \\
&\qquad \mathrm{Score}(p,v) \in [0,1].
\end{aligned}
\end{equation}

\begin{equation}
\begin{aligned}
\label{eq:maximum-weight-bipartite-matching}
\max_{M: P' \to V'} \quad 
  & \sum_{p \in P'} \mathrm{Score}\bigl(p, M(p)\bigr) \\
\text{subject to} \quad
  & \text{role}(p)  = \text{role}\!\bigl(M(p)\bigr), \\
  & \forall\, p \in P' .
\end{aligned}
\end{equation}

Allowing unrestricted cross-module matching substantially increases the number of candidate pairs, leading to structural ambiguity, degraded matching quality, and increased computational cost for large hierarchical designs. To address this issue, we enforce hierarchy-consistent alignment whenever corresponding module structures exist between the DUT and the golden model. Variables are preferentially matched within the same module scope, while cross-module mappings between sibling modules are disallowed. This constraint preserves structural consistency and prevents semantically unrelated variables from being incorrectly aligned. In practice, high-level golden models often adopt simplified structures that do not explicitly replicate the RTL hierarchy. Therefore, when a DUT submodule lacks a clear golden-model counterpart, we allow variables in the DUT top module to be matched against variables from arbitrary golden-model modules. Overall, the hierarchy-aware constraint significantly reduces the matching search space while preserving the flexibility required for heterogeneous cross-level designs. In our experiments, this strategy improves mapping stability and provides better scalability for complex hierarchical RTL designs.

\begin{figure*}[t]
    \centering
    \includegraphics[width=\linewidth]{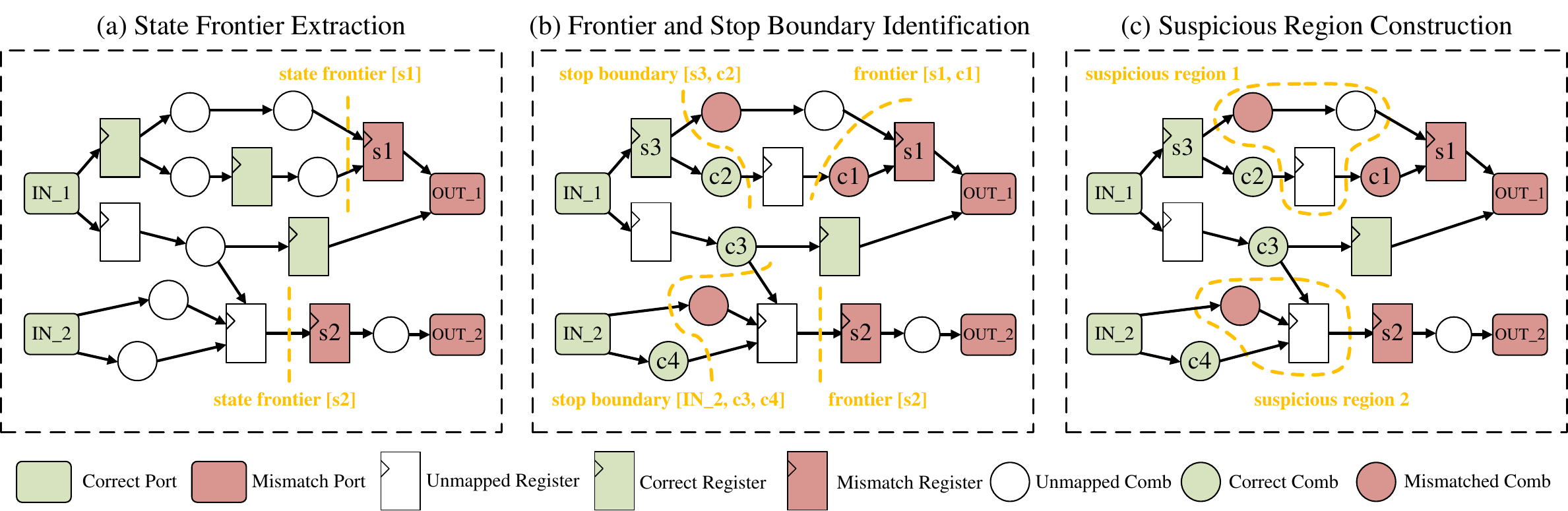}
    \caption{
    Progressive bug localization procedure in VeriPilot.
    (a) State frontier extraction identifies the earliest mismatched sequential states based on temporal and structural dependencies.
    (b) Frontier extraction performs guarded backtracing to identify the earliest upstream suspicious signals.
    (c) Suspicious region construction expands from frontier signals and collects all visited logic nodes until reaching stop boundaries.
    }
    \label{fig:BugLocation}
\end{figure*}

\subsection{Bug Localization} \label{sec:bug-localization}

In automated RTL debugging, merely identifying suspicious code lines is often insufficient to effectively bridge the gap between bug localization and bug repair. Isolated statements typically lack the structural and behavioral context required for LLM-based reasoning and patch generation. Instead of predicting individual buggy lines, our framework localizes bugs in the form of \emph{suspicious regions}, which provide compact yet semantically meaningful logic contexts surrounding the potential bug locations. To construct such regions, we propose a progressive localization procedure that jointly leverages temporal reasoning on sequential states and structural reasoning on combinational logic. The overall process incrementally refines the bug scope, starting from high-level behavioral divergence and gradually narrowing the analysis to compact structural logic cones, as illustrated in Fig.~\ref{fig:BugLocation}.

We first perform variable mapping on sequential state variables between the DUT and the golden model, and compare their runtime behaviors using execution traces. Among all mapped state variables, we identify the earliest mismatched states according to both temporal execution order and structural dependency. These variables are defined as the \emph{state frontier}, as illustrated in Fig.~\ref{fig:BugLocation}(a). Intuitively, the state frontier represents the earliest observable behavioral divergence between the DUT and the golden model at the sequential state level. Since sequential states provide stable temporal synchronization points between the two designs, anchoring localization on these states enables more reliable downstream structural analysis.

Starting from each state frontier element, we perform guarded backtracing through its structural predecessors in the constructed CDFG, as illustrated in Fig.~\ref{fig:BugLocation}(b). During traversal, we perform variable mapping on the involved combinational signals and classify them as mapped-and-matched, mapped-and-mismatched, or unmapped. If the predecessor combinational cone contains mapped-and-mismatched or unmapped signals, the corresponding state frontier together with these predecessor combinational signals are collected as the \emph{frontier}. Intuitively, the frontier represents the earliest structural boundary where the observed state-level divergence propagates from sequential states into combinational logic. The traversal terminates when reaching mapped-and-matched variables or primary inputs. Subsequently, starting from each frontier element, we further backtrace through its structural predecessors until all incoming signals become mapped-and-matched variables, primary inputs, or constants. Since these signals either behave consistently with the golden model or represent external design boundaries, they naturally terminate the traversal process. We denote these terminating signals as the \emph{stop boundary}, which defines the upstream limits of the suspicious logic region.

All sequential and combinational nodes visited during the above backtracing process are collected into the final \emph{suspicious region}, as illustrated in Fig.~\ref{fig:BugLocation}(c). Conceptually, this region corresponds to the minimal structural logic cone capable of explaining the observed behavioral divergence. Since the actual bug is expected to reside within or along the logic paths connecting the frontier and the stop boundary, the suspicious region provides a compact yet comprehensive search space for subsequent repair. Finally, suspicious regions derived from different frontier elements are merged whenever they are structurally connected. The remaining connected components are treated as independent bug candidates, enabling the framework to naturally support debugging scenarios involving multiple concurrent bugs.

\subsection{Bug Repair}
After bug localization, VeriPilot performs repair at the granularity of suspicious regions rather than isolated code lines. This design allows the repair process to preserve sufficient structural and behavioral context, which is critical for generating semantically correct RTL patches. For each suspicious region, we further leverage the frontier and stop-boundary signals identified during localization to retrieve their corresponding variables in the golden model through the proposed variable mapping framework.

Using these aligned variables as anchors, we extract the associated golden-model code segments that describe the expected behavior of the suspicious logic region. These reference snippets serve as semantically aligned repair guidance, enabling the LLM to reason not only about where the bug resides, but also about how the correct functionality should behave. Compared with relying solely on compiler errors or failing test cases, such cross-level behavioral guidance substantially improves the quality and stability of generated repairs.

Overall, VeriPilot summarizes the debugging results into structured debugging evidence consisting of four key components.
\begin{itemize}
    \item the \emph{frontier}, which captures the earliest structurally suspicious signals,
    \item the \emph{stop boundary}, which defines the upstream limits of the suspicious logic cone,
    \item the \emph{suspicious region}, which contains the potentially faulty RTL logic, and
    \item the aligned golden-model code segments, which describe the corresponding correct behavior.
\end{itemize}

Together, these artifacts provide both structural localization information and behavioral repair references for the LLM.

\begin{figure}[!t]
    \centering
    \includegraphics[width=1\linewidth]{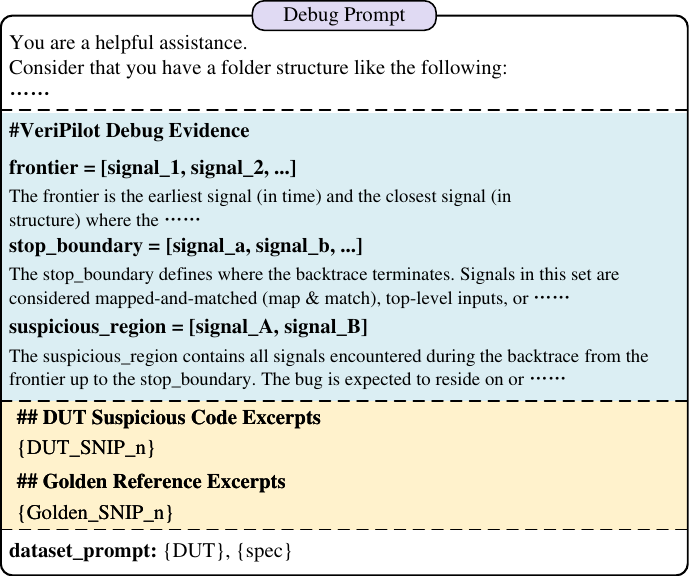}
    \caption{Bug repair prompt template used by VeriPilot, which integrates structured debugging evidence and relevant DUT/golden code snippets to guide LLM-based RTL repair.}
    \label{fig:DebugPrompt}
\end{figure}

Based on this information, we construct a tailored repair prompt, as illustrated in Fig.~\ref{fig:DebugPrompt}. The prompt begins with a system-level instruction that defines the debugging objective and repair constraints. Next, the structured debugging evidence generated by VeriPilot is provided, including the frontier signals, suspicious region, stop boundary, mapped variables, and the extracted golden-model reference code. Finally, the DUT source code and the design specification are appended to provide complete design semantics and implementation context.
Compared with directly feeding the entire RTL design into the LLM, this structured prompt significantly reduces irrelevant context and explicitly highlights the logic regions most relevant to the observed behavioral divergence. As a result, the LLM can focus its reasoning on the most suspicious structural dependencies and generate more accurate repair candidates.

The constructed prompt is then provided to the LLM to synthesize a repaired RTL patch. The generated patch is inserted back into the DUT, after which the updated design is re-evaluated using the testbench to regenerate execution traces and validate functional correctness. If mismatches still exist, VeriPilot re-enters the debugging stage and repeats the localization--repair loop until either a correct implementation is obtained or the maximum iteration limit is reached.
This iterative localization--repair workflow allows VeriPilot to progressively refine both bug understanding and repair quality across debugging iterations. By tightly coupling structured bug evidence with semantically aligned golden-reference guidance, the framework achieves substantially better repair accuracy and robustness than approaches that rely solely on test failures, compiler diagnostics, or unconstrained end-to-end code generation.

\section{Experiment Results}

\subsection{Experiment Setups}

\textbf{Software Setups:}
VeriPilot relies on several key software for the implementation. PyVerilog~\cite{takamaeda2015pyverilog} and pyslang~\cite{pyslang} are used to parse Verilog and SystemVerilog source files, extract AST and data-flow information, and construct the control-data-flow graph (CDFG) of the DUT. Since VeriPilot relies on golden model for the debugging, we extend the original verification harness with a base golden model abstraction to capture the evolution of internal variables during simulation. We implement executable Python-based golden models and leverage Python’s built-in AST utilities to analyze their semantics and construct the corresponding golden-side CDFGs. The DUT is compiled using a hardware simulator and accessed through cocotb’s VPI interface~\cite{cocotb}, which enables Python to control simulation execution, inject test stimuli, and collect both RTL signal traces and golden model variable traces. These components are further integrated with pytest to enable automated simulation, regression testing, and differential checking between the DUT and the golden reference. Furthermore, we employ gpt-3.5-turbo, gpt-4o and gpt-5 as the backbone LLMs for both DUT–golden variable mapping and automated patch generation.

\textbf{Evaluation Benchmark:} In this experiment, we evaluate our approach on both the Strider dataset and the CVDP benchmark. The Strider dataset, originally proposed for evaluating non-LLM debugging approaches, contains 57 real defect–fix pairs collected from seven programming assignments. In contrast, CVDP is a benchmark framework designed to evaluate LLM-based and agent-based solutions for hardware verification tasks, providing more complex and realistic debugging scenarios. This makes it well suited for evaluating the upper-bound capability and robustness of VeriPilot on challenging RTL designs. Compared with Strider, CVDP contains more diverse bug types, including missing ports, partially implemented modules, and other functional omissions. Moreover, these bugs are often distributed across multiple locations in the design rather than confined to a single defect pattern. In this work, we select the cid16 Design Verification - Debugging / Bug Fixing subset for evaluation. Since the original CVDP designs are written in SystemVerilog, while most prior RTL debugging approaches primarily target Verilog such as Cirfix and Strider, we manually translate these designs into semantically equivalent Verilog implementations to enable fair comparison with existing methods. Additionally, some CVDP prompts include repair hints that explicitly describe how the bug should be fixed. To more rigorously evaluate the bug reasoning capability of LLM-based approaches, we remove such repair guidance from the prompts to avoid leaking ground-truth repair. Basically, Strider represents a relatively simple benchmark and CVDP represents a more complex and diverse benchmark.

\subsection{Overall Debugging Capability Evaluation}

\begin{table}[t]
    \centering
    \caption{Repair performance across different benchmarks}

    \begin{tabular}{l c c c}
    \toprule
    
    \multirow{2}{*}{Type} 
    & \multirow{2}{*}{Model} 
    & \multicolumn{2}{c}{Benchmarks} \\
    
    \cmidrule(lr){3-4}
    & & Strider & CVDP-cid16 \\
    \midrule
    
    \multirow{2}{*}{Algorithm}
    & CirFix  & 15.79 & 22.85 \\
    & Strider & 36.84 & 28.57 \\
    \midrule
    \multirow[c]{3}{*}{LLM}
    & GPT-3.5-turbo & 43.85 & 34.29 \\
    & GPT-4o      & 78.95 & 57.14 \\
    & GPT-5       & 84.21 & 74.28 \\
    \midrule
    \multirow{2}{*}{RTL Agent}
    & MAGE(GPT-3.5-turbo) & 57.89  & 42.86 \\
    & MAGE(GPT-4o) & 87.72 & 65.71 \\
    & MAGE(GPT-5)  & 91.23 & 80.00 \\
    \midrule
    \multirow{2}{*}{Ours}
    & VeriPilot(GPT-3.5-turbo) & 70.18 & 51.42 \\
    & VeriPilot(GPT-4o) & 91.23 & 85.71 \\
    & VeriPilot(GPT-5)  & 92.98 & 91.43 \\
    \bottomrule
    \end{tabular}

    \label{tab:RepairResult}
\end{table}

Table~\ref{tab:RepairResult} summarizes the RTL repair performance of VeriPilot and representative baselines on the Strider and CVDP benchmarks. The compared approaches cover three categories: traditional algorithm-based debugging methods (CirFix and Strider), foundation LLMs with direct prompting, and the state-of-the-art RTL debugging agent MAGE. Across both benchmarks and all backbone models, VeriPilot consistently achieves the highest repair success rates, demonstrating the effectiveness of combining algorithmic debugging analysis with LLM-based repair generation.

Compared with traditional algorithm-based approaches, VeriPilot achieves substantially higher repair success rates on both benchmarks. On Strider, CirFix and Strider repair only 15.79\% and 36.84\% of defects, respectively, whereas VeriPilot repairs up to 92.98\% of buggy designs when equipped with GPT-5. Similar trends are observed on the more challenging CVDP benchmark, where VeriPilot reaches 91.43\% and outperforms CirFix and Strider considerably. The limited effectiveness of algorithm-based approaches primarily arises from their reliance on predefined repair templates and structural constraints. Although these methods are effective for addressing simple syntactic defects, they struggle with bugs in CVDP benchmark which generally has larger design contexts and requires deeper semantic reasoning about the intended functionality and coordinated modifications across multiple code regions.

Compared with direct LLM prompting, VeriPilot consistently delivers significant improvements across all backbone models. On Strider, repair rates increase from 43.85–84.21\% for standalone LLMs to 70.18–92.98\% with VeriPilot. Similar gains are observed on CVDP, where repair performance improves from 34.29–74.28\% to 51.42–91.43\%. These improvements indicate that model capability alone is insufficient for reliable RTL debugging. Although modern LLMs possess strong code understanding abilities, they still face difficulties in accurately identifying fault locations and inferring intended design behaviors from incomplete specifications. VeriPilot alleviates these challenges by providing semantically grounded debugging evidence, allowing the model to focus its reasoning on a substantially reduced search space.

MAGE is an agent-based baseline that enhances LLMs with iterative reasoning and tool-assisted debugging. Despite this, VeriPilot consistently outperforms MAGE across all backbone models and benchmarks. On Strider, VeriPilot raises the repair rate from 57.89\% to 70.18\% with GPT-3.5-turbo, and from 91.23\% to 92.98\% when moving from GPT-4o to GPT-5. On the more challenging CVDP benchmark, the improvements are even more pronounced, increasing repair rates from 42.86\%, 65.71\%, and 80.00\% to 51.42\%, 85.71\%, and 91.43\%, respectively. The key distinction is that VeriPilot explicitly leverages executable golden reference designs and structural analysis to generate debugging evidence before repair generation. This approach provides the LLM with more precise semantic guidance, resulting in higher repair accuracy and better scalability on complex RTL designs. The substantial performance gap on CVDP highlights that semantic alignment and structured debugging evidence become increasingly critical as design complexity grows.

In addition, an interesting observation is that the performance of VeriPilot remains strongly correlated with the capability of the underlying foundation model. When using GPT-3.5-turbo, VeriPilot achieves repair rates of 70.18\% and 51.42\% on Strider and CVDP, respectively. Replacing GPT-3.5-turbo with GPT-4o significantly improves the repair rates to 91.23\% and 85.71\%, indicating that stronger reasoning and code understanding capabilities directly translate into better repair outcomes. This observation suggests that VeriPilot does not replace the reasoning process of LLMs. Instead, it provides structured debugging evidence that allows the model to utilize its reasoning capacity more effectively. On the other hand, the improvement from GPT-4o to GPT-5 is considerably smaller. On Strider, the repair rate increases only from 91.23\% to 92.98\%, despite the approximately four-point gap between the two models when evaluated through direct prompting. A similar trend can be observed on CVDP. We believe this saturation effect may originate from benchmark limitations rather than insufficient model capability.

\subsection{VeriPilot Failure Analysis}
To better understand the limitations of VeriPilot, we analyze the failure cases when using GPT-4o and GPT-5. Table~\ref{tab} summarizes all failed designs. Despite the diversity of benchmark designs, the failures can be broadly categorized into two stages of the VeriPilot pipeline: (1) semantic alignment failures, where meaningful debugging evidence cannot be reliably extracted from the golden reference model, and (2) repair failures, where the fault is successfully localized but the LLM fails to synthesize a valid repair.

\begin{table}[t]
\centering
\caption{Categorization of remaining failure cases of VeriPilot with GPT-4o and GPT-5.}
\begin{tabular}{lll}
\toprule
Category & GPT-4o Failure Cases & GPT-5 Failure Cases \\
\midrule
Semantic misalignment 
& counter6bit (5 cases)
& counter6bit (4 cases) \\
\midrule
\makecell[l]{Localized but \\ difficult repair}
&
\makecell[l]{pipelined\_adder\_32bit\\
pipelined\_booth\_mult\\
montgomery\_mult\\
AXI\_ALU\\
galois\_encryption}
&
\makecell[l]{montgomery\_mult\\
AXI\_ALU\\
galois\_encryption}
\\
\bottomrule
\end{tabular}
\label{tab:failure_category}
\end{table}

\begin{figure}[htbp]
\centerline{\includegraphics[width=1\linewidth]{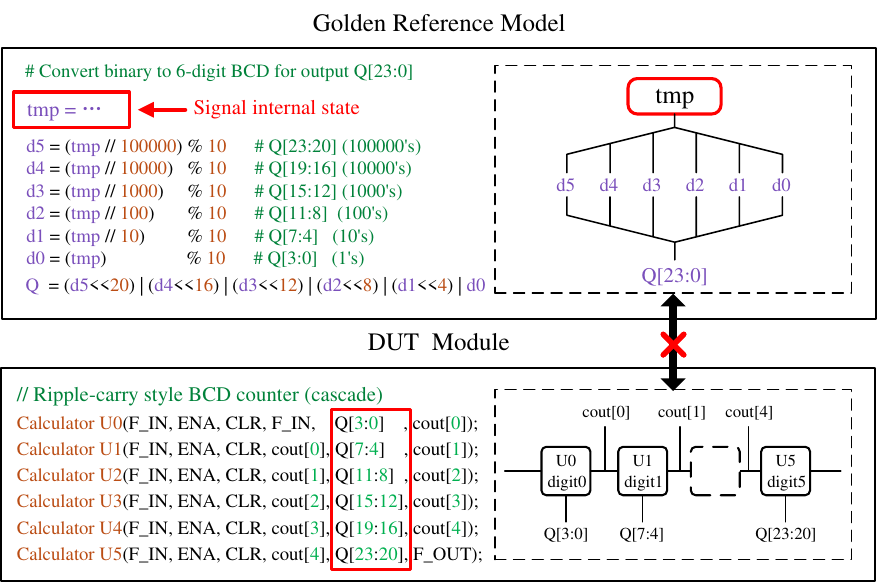}}
\caption{Semantic alignment failure in the counter6bit design. The golden model exposes a single binary state (tmp), while the DUT distributes the counting state across multiple internal registers, making reliable variable alignment difficult and potentially leading to misleading debugging evidence.}
\label{fig:challenge1}
\end{figure}

\textbf{Failure Type I: Semantic Alignment Failure.}
The first category originates from the semantic alignment stage. VeriPilot relies on the golden reference model to establish correspondences between DUT signals and reference variables. While the DUT and golden model may be functionally equivalent, successful alignment additionally requires sufficient correspondence between their internal state representations. A representative example is the \textit{counter6bit} benchmark shown in Fig.~\ref{fig:challenge1}. The golden reference maintains a single binary counter and derives the displayed BCD digits through arithmetic conversion. In contrast, the DUT directly stores each decimal digit as an independent state and propagates carries through cascaded calculator modules. Although both implementations satisfy the same functional specification, many internal DUT states have no meaningful counterparts in the golden model. Consequently, variable alignment becomes unreliable, and mismatched signals identified from the reference model may not accurately reflect the true root cause.

\begin{figure}[htbp]
\centerline{\includegraphics[width=1\linewidth]{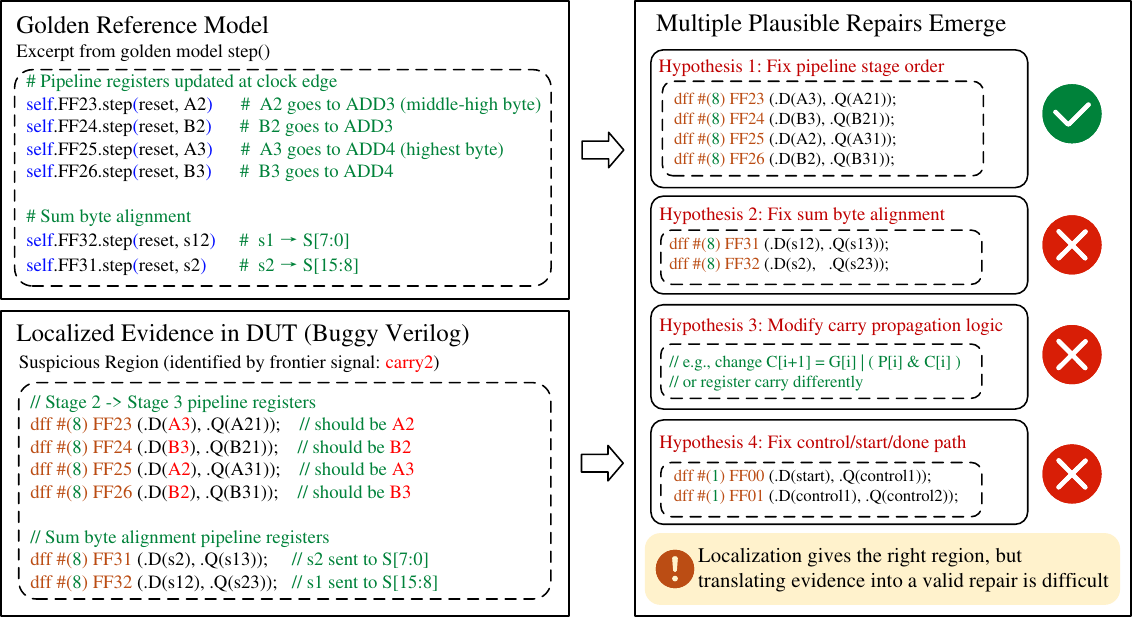}}
\caption{Repair ambiguity after successful bug localization. The localized suspicious region and golden-model evidence can lead to multiple plausible repair hypotheses, while identifying the correct patch requires deeper reasoning about pipeline interactions and temporal behavior.}
\label{fig:challenge2}
\end{figure}

\textbf{Failure Type II: Repair Construction Failure.}
The second category emerges after successful localization and becomes the dominant source of failures on the more complex CVDP benchmark. In these cases, VeriPilot is able to identify frontier signals, suspicious regions, and relevant golden-model evidence. However, transforming this debugging evidence into a correct implementation fix remains a challenging reasoning task.
The \textit{pipelined adder} example in Fig.~\ref{fig:challenge2} illustrates this challenge. VeriPilot accurately localizes the earliest mismatched signals and extracts relevant reference logic, but the suspicious region may admit multiple plausible repair hypotheses. Selecting the correct repair requires reasoning about temporal behavior, pipeline-stage interactions, and signal dependencies that are not explicitly encoded in the debugging evidence. Similar behavior can be observed in the modified Booth multiplier.

The impact of model reasoning capability becomes evident when comparing GPT-4o and GPT-5. Without changing the debugging framework, GPT-5 successfully repairs several previously unsolved designs, including the pipelined adder, modified Booth multiplier, and APB DSP operator. Since the generated debugging evidence remains largely unchanged, these improvements indicate that the bottleneck lies primarily in repair construction rather than fault localization. In other words, VeriPilot is often able to identify \emph{where} the problem occurs, while generating \emph{how} to fix it still depends on the reasoning capability of the backbone model.

Nevertheless, some failures persist even with GPT-5. These designs typically require architectural modifications rather than localized behavioral corrections. For example, repairing the Montgomery multiplier requires introducing additional submodules and inserting new pipeline stages, while the AXI ALU and Galois encryption designs involve large-scale missing functionality and multiple interacting defects. In such cases, the suspicious regions and aligned reference code may span a substantial portion of the design, leaving a large repair search space even after localization. Consequently, successful repair becomes increasingly constrained by the generative and architectural reasoning capabilities of the underlying model.

Overall, the failure analysis reveals a clear separation between fault localization and repair synthesis in RTL debugging. The remaining semantic-alignment failures indicate that accurate correspondence between heterogeneous implementations remains an open challenge. Meanwhile, the reduction in repair-construction failures from GPT-4o to GPT-5 suggests that VeriPilot already provides informative debugging evidence, but translating this evidence into correct repairs continues to depend heavily on the reasoning capabilities of the underlying LLM. These findings suggest that future RTL debugging frameworks should focus on both improving semantic alignment across diverse implementations and developing repair-oriented reasoning mechanisms that explicitly bridge the gap between debugging evidence and repair generation.

\subsection{Bug Localization Evaluation}

Since many benchmark designs are relatively small, we focus on the 16 most challenging repair cases that cannot be successfully repaired by GPT-4o alone to evaluate the effectiveness of VeriPilot's bug localization capability. Table~\ref{tab:bugloc_hard_cases} summarizes the localization results, including the total RTL lines of code (\textit{LOC}), the size of the localized suspicious region (\textit{Sus.}), and the resulting \textit{Search Space Reduction}, defined as

\begin{equation}
\textit{Search Space Reduction}
=1-\frac{\textit{Suspicious LOC}}{\textit{Total LOC}}.
\end{equation}

\begin{table}[t]
\centering
\caption{Bug localization quality on the 16 hard repair cases that GPT-4o fails without VeriPilot guidance.}
\label{tab:bugloc_hard_cases}
\begin{tabular}{lcccc}
\toprule
\textbf{Design} & \textbf{LOC} & \textbf{Sus.} & \textbf{Red.} & \textbf{Repair} \\
\midrule
pipelined\_adder\_32bit               & 131 & 19  & 85.5 & \xmark \\
arithmetic\_progression\_generator    & 75  & 27  & 64.0 & \cmark \\
caesar\_cipher                        & 63  & 50  & 20.6 & \cmark \\
coffee\_machine                       & 222 & 157 & 29.3 & \cmark \\
FILO\_RTL                             & 59  & 41  & 30.5 & \cmark \\
fsm\_seq\_detector                    & 125 & 19  & 84.8 & \cmark \\
image\_stego                          & 22  & 7   & 68.2 & \cmark \\
line\_buffer                          & 165 & 75  & 54.5 & \cmark \\
pipelined\_modified\_booth\_multiplier& 105 & 39  & 66.7 & \xmark \\
prim\_max\_find                       & 82  & 6   & 92.7 & \cmark \\
sobel\_filter                         & 60  & 11  & 81.7 & \cmark \\
apb\_dsp\_op                          & 165 & 19  & 88.5 & \cmark \\
axi\_alu                              & 132 & 51  & 61.4 & \xmark \\
galois\_encryption                    & 177 & 123 & 30.5 & \xmark \\
signed\_sequential\_booth\_multiplier & 141 & 112 & 20.6 & \cmark \\
montgomery\_mult                      & 151 & 26  & 82.8 & \xmark \\
\midrule
\textbf{Mean}   & 117 & 49 & 60.1 & 11/16 \\
\bottomrule
\end{tabular}
\end{table}

Overall, VeriPilot reduces the repair search space by an average of 58.4\% while preserving complete SystemVerilog code blocks as debugging evidence. Rather than returning isolated statements, VeriPilot localizes semantically coherent code regions, enabling the LLM to focus on relevant logic while retaining sufficient structural context for repair generation. The localization results demonstrate a strong correlation between search-space reduction and repair effectiveness. Among the 16 challenging designs, VeriPilot successfully repairs 11 cases after localization. In many successful repairs, the framework is able to narrow the debugging scope to a small fraction of the original design, substantially reducing the reasoning burden imposed on the LLM. These results suggest that accurate localization serves as an effective intermediate step between failure observation and repair generation. 

Interestingly, several unrepaired designs still exhibit high-quality localization results. For example, the 32-bit pipelined adder and the Montgomery multiplier achieve search-space reductions of 85.5\% and 82.8\%, respectively. In both cases, VeriPilot successfully isolates the relevant suspicious logic and identifies the earliest faulty signals. However, generating a correct repair requires coordinated modifications across multiple pipeline stages and non-local structural reasoning, which remains challenging even after the debugging scope has been significantly reduced. 

The effectiveness of localization also depends on the structural characteristics of the design. For example, the Galois encryption module contains signals with extensive fan-in and fan-out dependencies, causing the corresponding logic cone to span a large portion of the design. To preserve semantic completeness, VeriPilot includes the entire dependent code region as debugging evidence, resulting in a relatively smaller search-space reduction. Similar behavior can be observed in designs with highly interconnected control logic, where maintaining sufficient context is more important than aggressively minimizing the suspicious region.

Finally, some failures are fundamentally dominated by repair complexity rather than localization quality. The AXI ALU design requires synthesizing a substantial amount of missing functionality, expanding from approximately 132 RTL lines to more than 400 lines in the corrected implementation. Likewise, the Galois encryption design contains multiple distributed defects that require extensive modifications throughout the design. In such scenarios, localization alone provides limited benefit because the primary challenge lies in large-scale code generation rather than fault identification.

Overall, these results suggest that the primary role of VeriPilot is to reduce debugging uncertainty by narrowing the search space and generating structured debugging evidence. Even when repair generation ultimately fails, the localized suspicious regions significantly reduce the amount of RTL code requiring inspection, making root-cause analysis substantially more efficient. At the same time, the results highlight an inherent limitation of localization-based approaches: when defects correspond to missing functionality or require large-scale architectural modifications, accurate localization alone is insufficient, and repair success becomes increasingly dependent on the reasoning and synthesis capabilities of the underlying LLM.

\subsection{Variable Mapping Evaluation}

The effectiveness of VeriPilot's bug localization largely depends on the quality of variable mapping between the DUT and the golden model especially for more complex designs. Since localization is performed by comparing the runtime behavior of mapped variables, incorrect mappings may directly propagate to subsequent trace analysis and lead to inaccurate localization results. Therefore, we evaluate the proposed mapping strategy independently before analyzing its impact on bug localization. We select the 16 debugging tasks in the CVDP-cid16 benchmark that GPT-4o fails to solve for evaluation. For the baseline, we directly provide the DUT and golden model source code to GPT-4o and leverage it to generate variable correspondences in a predefined format, referred to as \textit{raw LLM mapping}. We then compare the generated mappings against VeriPilot's automated mapping strategy. The results are summarized in Table~\ref{tab:mapping}.

Overall, VeriPilot improves mapping precision from 55.89\% to 80.99\%, demonstrating the effectiveness of combining behavioral similarity with structural consistency constraints. Compared with raw LLM mapping, VeriPilot typically produces fewer candidate correspondences, but a substantially larger fraction of them are correct. This suggests that the proposed alignment framework effectively filters spurious matches while preserving semantically meaningful signal correspondences. The improvement is most pronounced in large and structurally complex designs, where variable correspondence cannot be reliably inferred from signal names alone. For example, the 32-bit pipelined adder contains hierarchical module instantiations and repeated pipeline stages, resulting in multiple signals with similar naming patterns and functionality. In such scenarios, direct LLM-based mapping frequently produces ambiguous correspondences. By incorporating runtime trace similarity, structural connectivity, and hierarchy-aware constraints, VeriPilot is able to distinguish signals that exhibit similar behavior but belong to different logical contexts, substantially improving mapping precision.

In contrast, the performance gap becomes much smaller in designs with limited mapping ambiguity. For instance, the FSM sequence detector contains only a small number of state variables, making most signal correspondences relatively straightforward to identify. As a result, both approaches achieve comparable mapping accuracy. Similar observations can be made for the image steganography encoder, where the design is dominated by combinational logic and the number of candidate variables remains relatively small. In these cases, the completeness of the golden model becomes a more important factor than the alignment algorithm itself. These results indicate that the primary benefit of VeriPilot lies in resolving ambiguity in complex RTL designs. As design size and structural complexity increase, behavioral and structural constraints become increasingly important for establishing reliable variable correspondences. The resulting improvement in mapping precision directly benefits subsequent mismatch detection and bug localization, where incorrect mappings can otherwise propagate errors throughout the debugging pipeline.

\begin{table}[t]
\centering
\caption{Signal alignment accuracy comparison between direct name matching and VeriPilot.
C/C denotes Candidate / Correct.}
\label{tab:mapping}
\begin{tabular}{lccc}
\toprule
\textbf{Benchmark} &
\textbf{Total} &
\textbf{Direct} &
\textbf{VeriPilot} \\
&
\textbf{Signals} &
\textbf{C/C} &
\textbf{C/C} \\
\midrule

pipelined\_adder\_32bit                 & 179 & 105 / 44 & 72 / 57 \\
arith\_progression\_gen                 & 18  & 14 / 7   & 8 / 8 \\
caesar\_cipher                          & 13  & 11 / 7   & 9 / 8 \\
coffee\_machine                         & 25  & 18 / 8   & 12 / 10 \\
FILO\_RTL                               & 37  & 22 / 9   & 15 / 12 \\
fsm\_seq\_detector                      & 11  & 11 / 8   & 9 / 8 \\
image\_stego                            & 9   & 6 / 5    & 5 / 5 \\
line\_buffer                            & 110 & 74 / 41  & 60 / 53 \\
pipelined\_booth\_mult                  & 23  & 24 / 9   & 19 / 11 \\
prim\_max\_find                         & 84  & 54 / 35  & 45 / 39 \\
sobel\_filter                           & 18  & 14 / 4   & 10 / 5 \\
apb\_dsp\_op                            & 90  & 41 / 29  & 48 / 44 \\
axi\_alu                                & 40  & 18 / 13  & 21 / 19 \\
galois\_encryption                      & 314 & 118 / 80 & 102 / 71 \\
seq\_signed\_booth\_mult                & 22  & 13 / 6   & 10 / 8 \\
montgomery\_mult                        & 40  & 26 / 13  & 18 / 17 \\

\midrule
\textbf{Overall Precision} & -- &
\textbf{55.89\%} &
\textbf{80.99\%} \\
\bottomrule
\end{tabular}

\end{table}

\subsection{Token Consumption Evaluation}

Due to the inherent nondeterminism of LLM-based repair generation, a design may require multiple repair attempts before a correct solution is obtained. To eliminate the impact of varying iteration counts, we report the average token consumption per repair iteration. The results are summarized in Table~\ref{tab:token_cost}. Compared with raw prompting, VeriPilot increases token consumption by an average of 2.58$\times$. This overhead mainly originates from two sources. First, additional LLM queries are required during the variable-alignment stage to establish semantic correspondences between DUT and golden-model signals. Second, VeriPilot augments the repair prompt with structured debugging evidence, including variable mappings, frontier signals, suspicious regions, and relevant golden-reference code fragments. The additional token usage grows with design complexity. Large designs such as the coffee-machine controller, AXI ALU, and Galois encryption module require more extensive debugging evidence and therefore incur higher token costs, whereas smaller designs introduce only modest overhead. This trend is expected, as larger designs contain more candidate signals and broader suspicious regions that must be analyzed and communicated to the LLM. Despite the increased token consumption, the additional context substantially improves repair effectiveness by enabling the LLM to reason about the internal behavior of the design. Overall, the results demonstrate a favorable trade-off between token cost and repair performance: VeriPilot incurs moderate overhead while providing significantly stronger localization capability and higher repair success rates.

\begin{table}[t]
\centering
\caption{Token consumption comparison between raw prompting and VeriPilot.}
\label{tab:token_cost}
\begin{tabular}{lcc}
\toprule
Design & Raw Tokens & VeriPilot Tokens \\
\midrule
pipelined\_adder\_32bit      & 4791 & 16195 \\
arith\_progression\_gen      & 1743 & 4982 \\
caesar\_cipher               & 1557 & 5063 \\
coffee\_machine              & 8152 & 25476 \\
FILO\_RTL                    & 1603 & 3535 \\
fsm\_seq\_detector           & 1787 & 8485 \\
image\_stego                 & 1297 & 2902 \\
line\_buffer                 & 6165 & 14023 \\
pipelined\_booth\_mult       & 2902 & 6485 \\
prim\_max\_find              & 2381 & 4465 \\
sobel\_filter                & 5738 & 8234 \\
string\_to\_ascii            & 5385 & 7727 \\
apb\_dsp\_op                 & 5694 & 12217 \\
axi\_alu                     & 8999 & 19108 \\
galois\_encryption           & 9211 & 22856 \\
seq\_signed\_booth\_mult     & 2079 & 3814 \\
montgomery\_mult             & 5432 & 10490 \\
\midrule
Average Overhead             & \multicolumn{2}{c}{2.58$\times$} \\
\bottomrule
\end{tabular}
\end{table}

\subsection{Ablation Study}

To evaluate the contribution of different forms of structured debugging evidence, we conduct an ablation study across both Strider and CVDP-cid16 using three backbone models: GPT-3.5-turbo, GPT-4o, and GPT-5. We progressively augment the repair prompt with different evidence sources, including suspicious regions (\textbf{B}) and golden-model references (\textbf{G}), starting from the baseline setting that contains only the specification and mismatch information (\textbf{S}). This allows us to isolate the contribution of each evidence type and analyze how their effectiveness varies across models of different capabilities. The results are summarized in Table~\ref{tab:ablation}.

On the Strider dataset, providing the ground-truth bug location significantly improves repair accuracy. This improvement can be attributed to two factors. First, it eliminates the need for the LLM to perform additional reasoning for bug localization. Second, it constrains the modification scope, preventing the LLM from unnecessarily altering correct logic or attempting to rewrite large portions of the code, which often introduces additional errors. Providing the golden model further improves repair performance beyond supplying bug locations alone. This is because the comparison between the DUT and the golden reference implicitly encodes bug localization information while also providing guidance on how the faulty behavior should be corrected, thereby facilitating more effective repair generation.

On the CVDP dataset, a similar trend is observed. Moreover, the performance gain of {S+G} over {S+B} is more pronounced than that on the Strider dataset. This can be attributed to two factors. First, the use of stronger LLM backbones amplifies the benefit of structured debugging evidence. Second, the higher implementation diversity in Strider introduces a larger abstraction gap between DUTs and golden models, which reduces the effectiveness of golden-model guidance compared to the more structurally consistent designs in CVDP-cid16.

\begin{table}[t]
\centering
\caption{Ablation study of evidence used in LLM-based repair.}
\begin{tabular}{llcccc}
\toprule
Benchmark & Model & S & S+B & S+G & S+B+G \\
\midrule

\multirow{3}{*}{Strider}
& GPT-3.5-turbo & 43.85 & 56.14 & 64.91 & 70.18 \\
& GPT-4o        & 78.95 & 84.21 & 89.47 & 91.23 \\
& GPT-5         & 84.21 & 87.72 & 91.23 & 92.98 \\
\midrule

\multirow{3}{*}{CVDP-cid16}
& GPT-3.5-turbo & 34.29 & 42.86 & 48.57 & 51.43 \\
& GPT-4o        & 57.14 & 68.57 & 80.00 & 85.71 \\
& GPT-5         & 74.28 & 82.86 & 88.57 & 91.43 \\
\bottomrule
\end{tabular}
\label{tab:ablation}
\end{table}

\section{Conclusion}

We presented VeriPilot, a unified framework that combines static structural analysis, dynamic simulation evidence, and LLM-driven reasoning for automated Verilog debugging and repair. By exposing fine-grained internal behaviors through CDFGs, semantic variable mapping, frontier-signal extraction, and guarded backtracing, VeriPilot provides LLMs with precise and structured debugging evidence, enabling them to reason about root causes rather than relying on unguided code regeneration.
Experimental results on both Strider and CVDP-cid16 demonstrate the effectiveness of the proposed framework. When integrated with GPT-4o, VeriPilot achieves repair rates of 91.23\% and 85.71\% on Strider and CVDP-cid16, respectively. With GPT-5, the repair rates further increase to 92.98\% and 91.43\%, substantially outperforming traditional automated program repair approaches, standalone LLMs, and existing RTL debugging agents. 
These results suggest that combining program-analysis techniques with LLM reasoning is a promising direction toward scalable and reliable AI-assisted hardware development.

\section{Acknowledgments}
The authors used GPT-5 to polish the manuscript’s text. All core ideas, methodologies, and scientific conclusions are entirely the original work of the authors, who have thoroughly reviewed all AI-assisted outputs and take full responsibility for the final content.

\newpage

\bibliographystyle{unsrt}
\bibliography{reference}

\end{document}